# Spontaneous symmetry breaking as a resource for noncritically squeezed light

G. J. de Valcárcel[a†], F. V. Garcia-Ferrer[a], R. M. Höppner[a], I. Pérez-Arjona[b],
C. Navarrete-Benlloch[a], and E. Roldán[a‡]

[a]Departament d'Òptica, Universitat de València, Dr. Moliner 50, 46100-Burjassot, Spain.

[b]IGIC, Universidad Politécnica de Valencia, C/ Paranimf 1, 46730-Grao de Gandia, Spain.

[†] german.valcarcel@uv.es; phone +34 963 544 079; fax +34 963544 715.

[‡] eugenio.roldan@uv.es; phone +34 963 543 806; fax +34 963 544 715.

## ABSTRACT

In the last years we have proposed the use of the mechanism of spontaneous symmetry breaking with the purpose of generating perfect quadrature squeezing. Here we review previous work dealing with spatial (translational and rotational) symmetries, both on optical parametric oscillators and four-wave mixing cavities, as well as present new results. We then extend the phenomenon to the polarization state of the signal field, hence introducing spontaneous polarization symmetry breaking. Finally we propose a Jaynes-Cummings model in which the phenomenon can be investigated at the single-photon-pair level in a non-dissipative case, with the purpose of understanding it from a most fundamental point of view.

## 1. INTRODUCTION

### 1.1 Motivation

Non-classical states of light have been a subject of active research in the last decades. Squeezed states are probably one of the most simple examples of them: in contrast to the coherent states (like the usual *laser* light or the *vacuum*), where both quadratures of light have the same uncertainty, in an ideal squeezed state quantum fluctuations are reordered such as one of the quadratures is free from noise, while the other is completely undetermined[1].

Soon after the concept of squeezed light was introduced, it was proved that nonlinear resonators were able to generate it[1]. In particular, it was shown that both degenerate optical parametric oscillators (DOPOs) and degenerate four-wave mixing cavities –which are, respectively, optical resonators with a $\chi^{(2)}$ nonlinear crystal and a $\chi^{(3)}$ nonlinear medium inside–, were able to create a highly squeezed vacuum in the output field at the degenerate frequency (signal frequency in the following) when working close to their emission threshold. However, DOPOs operated below threshold are nowadays the most common source for squeezed light, most of all because $\chi^{(3)}$ media are usually affected by residual processes that provide extra noise (such as spontaneous emission or different types of scattering). Although noise reduction cannot be complete, as this would entail infinite fluctuations in the anti-squeezed quadrature (which requires infinite energy), squeezing levels as large as 11.5 dB (more than 90% of noise reduction) have been proved[2]. On the other hand, the squeezing level attained at threshold degrades as the system is brought apart from it, and hence this squeezing is *critical* as it requires a tuning of the system parameters.

Squeezed light has found major applications in several fields like high precision measurements[3,4] and quantum information[5], a reason why it is important to keep improving its quality and finding new sources able to generate it. This was the main motivation that led us to propose the study of the phenomenon of spontaneous symmetry breaking (SSB) in nonlinear cavities as a potential resource for squeezing.

### 1.2 General description of the phenomenon: squeezing induced by spontaneous symmetry breaking

The basic ideas behind this phenomenon can be put as follows. Suppose that the nonlinear cavity showing bifurcation squeezing is also invariant under changes of some continuous degree of freedom ε of the signal field (like, e.g., the orientation of its linear polarization), which we might call the *free parameter* (FP) in the following. Above threshold, the classical or mean field value of the signal field $\overline{\mathbf{E}}_\varepsilon(\mathbf{r},t)$ is not zero (we explicitly denote the FP in the field), and when the system starts emitting it, a particular value of the FP must be chosen (selected randomly according to the initial fluctuations), hence breaking the symmetry of the system. This is what we mean by SSB.

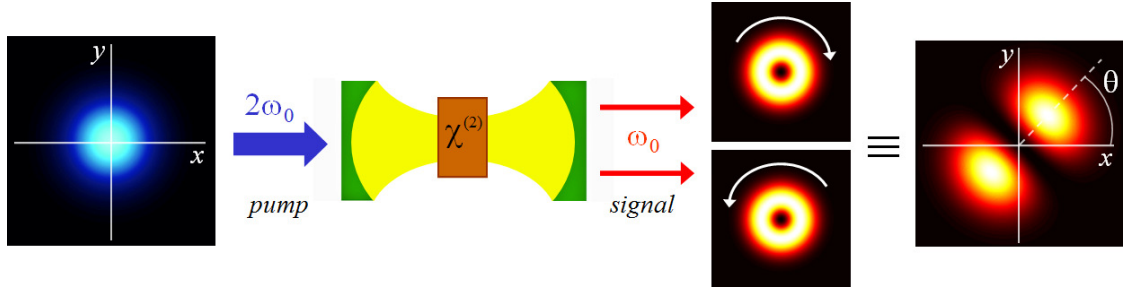

Figure 1. Scheme of the two-transverse-mode DOPO. The cavity is pumped by a Gaussian $TEM_{00}$ mode, and is tuned in such a way that the down-converted photons must be generated in opposite OAM pairs with an arbitrary phase difference. On the other hand, the superposition of two LG beams is equivalent to a HG $TEM_{10}$ mode whose orientation is given by the phase difference between the subjacent LG modes. Hence, above threshold mean field emission will take place in a $TEM_{10}$ mode with an arbitrary orientation.

Now think about what quantum theory might bring. As variations in the FP do not affect the system, quantum fluctuations will be able to make it fluctuate without opposition, eventually making it become completely undetermined. But invoking now the uncertainty principle, the complete indetermination of a system's variable allows for the perfect determination of its corresponding momentum. This means that we could expect squeezing to appear in the mode $-i\partial_\varepsilon \overline{\mathbf{E}}_\varepsilon(\mathbf{r},t)$. Moreover, this squeezing could be perfect (as the FP can be completely undetermined in the long term), and noncritical, as it is a product of the SSB which happens for any value of the parameters above threshold.

In this review we show that this intuitive result is correct by focusing on three particular symmetries: rotational and translational symmetries in the transverse plane (Section 2), and symmetry in a polarization parameter of the signal field (Section 3). In addition we will have the opportunity to study the process at the very fundamental level by using a single-atom analog of the polarization symmetry breaking which we model by a Jaynes-Cummings model (Section 4).

## 2. SPATIAL SYMMETRY BREAKING

### 2.1 Introduction

As a first example of squeezing induced by SSB we consider nonlinear resonators which have some spatial symmetry in the transverse plane. Although the first symmetry we studied was the translational symmetry in wide aperture DOPOs[6,7], we prefer to introduce here the phenomenon by using the spontaneous rotational symmetry breaking that occurs in two-transverse-mode DOPOs[8,9]. This system will allow us to expose the basic mathematical formalism that we use to analyze SSB from a quantum viewpoint, and to introduce important considerations about its experimental realization and the measurement of its related quadrature squeezing.

### 2.2 Spontaneous rotational symmetry breaking in two-transverse-mode DOPOs

Consider a cavity having a thin $\chi^{(2)}$ crystal placed on its waist plane and tuned in the following way[8,9,10,11,12]: At some frequency $2\omega_0$ (pump frequency) the cavity is tuned to a $TEM_{00}$ Gaussian mode $G(\mathbf{r})$, while at frequency $\omega_0$ (signal frequency) it is the first family of transverse modes that resonates.

The first family of transverse modes is formed by two Laguerre-Gauss (LG) modes $L_{\pm 1}(\mathbf{r})$ with $\pm 1$ orbital angular momentum (OAM), and this is the reason why we will call this system the two-transverse-mode DOPO. The explicit expression of these modes at the waist plane is given by[13]

$$G(\mathbf{r}) = \sqrt{2/\pi}\, w_p^{-1} \exp(-r^2/w_p^2) \quad , \quad L_{\pm 1}(\mathbf{r}) = 2\sqrt{\pi}\, r w_s^{-2} \exp(-r^2/w_s^2) e^{\pm i\phi}, \tag{1}$$

where $w_{p/s}$ is the beam radius at the pump/signal frequency, and $\mathbf{r} = r(\cos\phi, \sin\phi)$ is the coordinate vector in the transverse plane. Note that from the LG modes one can obtain the usual $TEM_{10}$ Hermite-Gauss (HG) mode rotated an angle $\psi$ respect to the horizontal as[13]

$$H_{10}^\psi(\mathbf{r}) = [e^{-i\psi} L_{+1}(\mathbf{r}) + e^{i\psi} L_{-1}(\mathbf{r})]/\sqrt{2} = \sqrt{2}|L_{\pm 1}(\mathbf{r})|\cos(\phi - \psi). \tag{2}$$

The Hamiltonian of the system has two parts (see Fig. 1): the pumping of the coherent, Gaussian, resonant mode at frequency $2\omega_0$, and the parametric down conversion of the pump photons into signal photons inside the crystal. In the interaction picture it reads

$$\hat{H} = i\hbar(\mathcal{E}_p \hat{a}_0^\dagger + \chi \hat{a}_0 \hat{a}_{+1}^\dagger \hat{a}_{-1}^\dagger) + \text{H.c.}, \tag{3}$$

where $\hat{a}_j^\dagger$ is the creation operator for a pump photon ($j = 0$) and signal photons with ±1 OAM ($j = \pm 1$), and $\mathcal{E}_p$ and $\chi$ are proportional to the amplitude of the pumping beam and the nonlinear susceptibility of the $\chi^{(2)}$ crystal, respectively[9]. The down-conversion part of this Hamiltonian is justified by energy and OAM conservation: from one $2\omega_0$ photon with zero OAM, two $\omega_0$ photons with opposite OAM are created.

This Hamiltonian has the symmetry $(\hat{a}_{+1}, \hat{a}_{-1}) \to (e^{i\theta}\hat{a}_{+1}, e^{-i\theta}\hat{a}_{-1})$, which leaves the phase difference $2\theta$ between the LG modes undefined. Hence $\theta$ is the FP of this system. The constant of motion associated to this symmetry is the photon number difference $\hat{a}_{+1}^\dagger \hat{a}_{+1} - \hat{a}_{-1}^\dagger \hat{a}_{-1}$. This ensures that the LG modes $L_{\pm 1}(\mathbf{r})$ will be twin beams whose intensity difference is potentially perfectly squeezed. Hence, once the threshold for signal's mode generation is crossed, mean-field emission will take place in the mode $H_{10}^\theta(\mathbf{r}) = [e^{-i\theta}L_{+1}(\mathbf{r}) + e^{i\theta}L_{-1}(\mathbf{r})]/\sqrt{2}$, that is, a HG mode whose dipole-like pattern breaks the rotational symmetry of the system and can appear classically along any orientation $\theta$. Using the techniques we explain in the next subsection, it can be shown that quantum noise makes rotate the pattern randomly, and that the phase quadrature of the HG mode orthogonal to the generated one is perfectly squeezed irrespective of the system parameters –which makes sense by following the reasoning given in the introduction, as $i\partial_\theta H_{10}^\theta(\mathbf{r}) = iH_{10}^{\theta+\pi/2}(\mathbf{r})$–. In the following we name $H_{10}^\theta(\mathbf{r})$ the *bright mode* (as it is classically excited), and its orthogonal mode $H_{10}^{\theta+\pi/2}(\mathbf{r})$ the *dark mode* (as it is classically empty of photons), denoting them by the indices '*b*' and '*d*' respectively when needed.

## 2.3 The mathematical formalism

Let us briefly explain the techniques we use to analyze the phenomenon of SSB from a quantum viewpoint, by using the two-transverse-mode DOPO as an example.

Hamiltonian (3) does not recast all the processes which occur in the DOPO. In particular it does not account for the photons which are leaving the cavity through the partially reflecting mirror; this is not a reversible process, and we must introduce it in the master equation satisfied by the density operator of the system[14]. We use a positive *P* representation for the density matrix[15] which allows us to map the master equation into the following stochastic (Langevin) equations[9]

$$\dot{\alpha}_0 = \mathcal{E}_p - \gamma_p \alpha_0 - \chi \alpha_{+1}\alpha_{-1}, \qquad \dot{\alpha}_0^+ = \mathcal{E}_p - \gamma_p \alpha_0^+ - \chi \alpha_{+1}^+ \alpha_{-1}^+,$$
$$\dot{\alpha}_{+1} = -\gamma_s \alpha_{+1} + \chi \alpha_0 \alpha_{-1}^+ + \sqrt{\chi \alpha_0}\xi(t), \qquad \dot{\alpha}_{+1}^+ = -\gamma_s \alpha_{+1}^+ + \chi \alpha_0^+ \alpha_{-1} + \sqrt{\chi \alpha_0^+}\xi^+(t), \qquad (4)$$
$$\dot{\alpha}_{-1} = -\gamma_s \alpha_{-1} + \chi \alpha_0 \alpha_{+1}^+ + \sqrt{\chi \alpha_0}\xi^*(t), \qquad \dot{\alpha}_{-1}^+ = -\gamma_s \alpha_{-1}^+ + \chi \alpha_0^+ \alpha_{+1} + \sqrt{\chi \alpha_0^+}[\xi^+(t)]^*,$$

where $\gamma_{p/s}$ is the cavity damping rate at the pump/signal frequency, and $\xi(t)$ and $\xi^+(t)$ are independent complex noises satisfying the usual complex white noise statistics[9].

The equivalence between these stochastic equations and the master equation must be understood in the following way: $\langle :f(\hat{a}_j, \hat{a}_j^\dagger): \rangle = \langle f(\alpha_j, \alpha_j^+) \rangle_{\text{stochastic}}$, that is, quantum averages of an operator function in normal order equal stochastic averages of the same function changing the boson operators by their associated stochastic variables.

Outside the cavity there exists a continuum of modes, and hence squeezing cannot be defined via the simple single-mode uncertainty. As was first shown by Collet and Gardiner[16], the quantity accounting for the fluctuations of quadrature $\hat{X}_m^\varphi = e^{-i\varphi}\hat{a}_m + e^{i\varphi}\hat{a}_m^\dagger$ outside the cavity (*m* refers to any signal transverse or polarization mode of the DOPO) is

$$V(\omega; X_m^\varphi) = 1 + 2\gamma_m \int d\tau \langle :\delta\hat{X}_m^\varphi(t)\delta\hat{X}_m^\varphi(t+\tau): \rangle e^{-i\omega\tau}, \qquad (5)$$

where $\delta A = A - \langle A \rangle$. We shall call this the *noise spectrum*, and $\omega$ the *noise frequency*. For the vacuum or coherent state we have $V = 1$; hence if $V < 1$ one can state that light is in a squeezed state for mode *m*. Note that the j = 0 (p/2) quadrature is usually called the *amplitude* (*phase*) *quadrature*, and is denoted by $X_m$ ($Y_m$).

Note that we can retrieve the classical equations from the quantum Langevin equations by setting the noises to zero and making $\alpha_j^+ = \alpha_j^*$. Then a simple algebraic manipulation, plus a stability analysis, shows that for $\mathcal{E}_p > \mathcal{E}_{\text{th}} = \gamma_p\gamma_s/\chi$ the trivial solution $\bar{\alpha}_{\pm 1} = 0$ becomes unstable, and a new stable solution $\bar{\alpha}_{\pm 1} = \rho\exp(\mp i\theta)$, with $\rho^2 = (\mathcal{E}_p - \mathcal{E}_{\text{th}})/\chi$, appears. Any value of $\theta$ is allowed by the classical emission, as the mean field equations preserve the symmetry of the original Hamiltonian. Note that this solution gives raise to the bright mode $H_{10}^\theta(\mathbf{r})$ as commented above.

As for the quantum dynamics, we have two methods[9] for solving Eqs. (4). The first one consist in assuming that quantum fluctuations are small as compared with the mean field solution, and then linearize Eqs. (4) with respect to them. When working with SSB we have to take special care because we expect $\theta$ to have arbitrarily large fluctuations. This means that linearization is only possible if the signal variables are expanded as

$$\alpha_{\pm 1} = [\rho + b_{\pm 1}(t)]e^{\mp i\theta(t)} \text{ and } \alpha_{\pm 1}^+ = [\rho + b_{\pm 1}^+(t)]e^{\pm i\theta(t)}, \quad (6)$$

as in this way $\theta$ carries with the larger part of the fluctuations, while the $b$'s and $\dot{\theta}$ remain small[8,9]. This expansion allows us to further track the evolution of the orientation of the generated pattern. The linearization leads to the linear system[8,9]

$$-2i\rho\mathbf{w}_0\dot{\theta} + \dot{\mathbf{b}} = \mathcal{L}\mathbf{b} + \sqrt{\gamma_s}\xi(t), \quad (7)$$

where $\mathbf{b}$ and $\xi$ are vectors collecting the fluctuations and the noise sources. $\mathcal{L}$ is a matrix whose particular expression is not important[8,9]; what is important is that it possesses two eigenvectors $\mathbf{w}_0$ and $\mathbf{w}_1$, having eigenvalues 0 and $-2\gamma_s$, respectively, which are related to the quadratures of the dark mode by $X_d = 2^{1/2}i\mathbf{w}_0\cdot\mathbf{b}$ and $Y_d = 2^{1/2}i\mathbf{w}_1\cdot\mathbf{b}$. After projecting the linear system onto $\mathbf{w}_0$ (we can further take $\mathbf{w}_0\cdot\mathbf{b} = 0$ as it just entails a redefinition of the arbitrary phase $\theta$) we get

$$\dot{\theta} = \frac{\sqrt{\gamma_s}}{2\rho}\eta_0(t) \Rightarrow V_\theta = \langle\delta\theta^2(t)\rangle_{\text{stochastic}} = \frac{d}{\sigma-1}\gamma_s t,, \quad (8.1)$$

$$\dot{Y}_d = -2\gamma_s Y_d + i\sqrt{2\gamma_s}\eta_1(t) \Rightarrow V(Y_d;\omega) = (\omega/2\gamma_s)^2 / \left[1 + (\omega/2\gamma_s)^2\right], \quad (8.2)$$

where $\eta_0$ and $\eta_1$ are real, independent noises with white noise statistics, $d = \chi^2/4\gamma_p\gamma_s$, and $\sigma = \mathcal{E}_p/\mathcal{E}_{\text{th}}$.

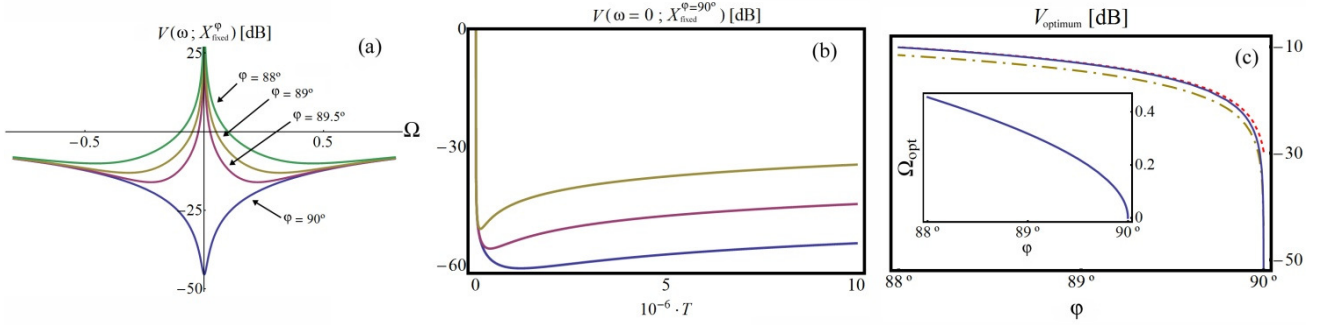

Figure 2. Noise spectrum as measured by a fixed TEM$_{10}$ LOF initially orthogonal to the bright mode. Results are given in dB, defined through $V[\text{dB}] = 10 \log V$. (a) Noise spectrum as a function of the noise frequency, and for different values of the LOF phase $\varphi$. The rest of parameters are $T = T_{\text{opt}}$, $d = 10^{-10}$, $\sigma = 2^{1/2}$ (this value for $\sigma$ is used also for (b) and (c), as the results are weakly dependent of it above threshold), although the same qualitative behavior is found for other values of them. (b) Noise spectrum at zero noise frequency and for $\varphi = 90°$ as a function of the detection time. Three different values of $d$ are considered ($10^{-11}$, $10^{-12}$, and $10^{-13}$ from top to bottom). Note the existence of an optimum detection time. (c) Noise spectrum evaluated for the optimum detection time and frequency as a function of the LOF phase ($d = 10^{-13}$ for the blue solid-line, and $d = 10^{-6}$ for the red-dashed one). The inset shows the optimum noise frequency as a function of $\varphi$ (which is independent of the rest of parameters).

The first expression shows that the orientation of the bright mode (the FP of this particular system) evolves ruled by quantum noise, thus becoming completely undetermined in the long term. Note that the random rotation of the pattern is slow if we work far enough from threshold ($d \approx 10^{-13}$ for common system parameters[8,9]). On the other hand, the second expression states that perfect squeezing appears at zero noise frequency in the phase quadrature of the dark mode, irrespective of the system's parameters.

The second method we used to solve equations (4) was a semi-implicit numerical algorithm first developed by Drummond and Mortimer[17], with which we proved that the analytical expressions obtained within the linear approximation are correct outside that limit[9].

## 2.4 Considerations about the measurement

The noise spectrum of a given quadrature of the system is measured with a homodyne detection scheme. It consists in mixing the light exiting the DOPO with an intense local oscillator field (LOF) in a beam splitter, subtracting the photocurrents measured on its output ports, and introducing the resulting signal in a spectrum analyzer. It is possible to show that when the LOF is prepared in mode $m$ and with phase $\varphi$, this measurement scheme directly gives the noise spectrum of quadrature $\hat{X}_m^\varphi$.

Therefore, in order to measure the squeezing properties of the dark mode, the LOF must be prepared in the mode orthogonal to the bright one, what seems impossible because this mode is rotating randomly. This is the reason why we studied the levels of squeezing that can be achieved when the local oscillator is prepared in the mode orthogonal to the bright one at the beginning of the experiment, but is kept fixed during the time the measurement lasts[9].

Let us call $T$ and $\Omega$ the adimensional detection time and the noise frequency normalized to $\gamma_s$. An approximate expression (valid in the limit of small $d$) for the noise spectrum was found to be[9,18]

$$V(\omega;\hat{X}^{\varphi}_{\text{fixed}}) = 1 + S^0(\omega)\cos^2\varphi + S^{\pi/2}(\omega)\sin^2\varphi, \qquad (9.1)$$

$$S^0(\omega) = \frac{8}{\omega^2}(1-\text{sinc}\,\Omega T) - \frac{4dT}{\Omega^2(\sigma-1)} \times \frac{6(\sigma-1)^2+\Omega^2}{4(\sigma-1)^2+\Omega^2}, \qquad (9.2)$$

$$S^{\pi/2}(\omega) = \frac{8-2\Omega^2}{T(4+\Omega^2)^2} - \frac{4}{4+\Omega^2} + \frac{8dT}{\Omega^2(\sigma-1)} \times \frac{2(\sigma^2+1)+\Omega^2}{(\sigma-1)(4+\Omega^2)(4\sigma^2+\Omega^2)}. \qquad (9.3)$$

In Fig. 2a we show this noise spectrum as a function of the $\Omega$ for different values of $\varphi$ and fixed values of the rest of parameters. It can be appreciated that the best squeezing levels are found for $\varphi = 90°$ and at zero noise frequency. In Fig. 2b we show the noise spectrum for this parameters as a function of $T$ and for different values of $d$ ($\sigma$ is fixed to $2^{1/2}$ as it can be checked that the results are almost independent of it above threshold). We see that there exists an optimum detection time $T_{opt} = [\sigma^2(\sigma-1)/d(\sigma^2+1)]^{1/2}$ which maximizes the squeezing. Nevertheless, for any value of the detection time above $T_{opt}$ the squeezing levels are quite large. On the other hand, in experiments[19,20] it is difficult to ensure that $\varphi = 90°$ above a precision of 1.5°. This is the reason why we have analyzed the noise spectrum for different values of it. Note that the optimum noise frequency is no longer $\omega = 0$, because the infinite fluctuations of $S^0$ at zero noise frequency enter the spectrum (Fig. 2a). On the other hand, it is possible to show that the optimum detection time is almost independent of $\varphi$ for small deviations of this from 90°, and hence it is still given to a good approximation by the previous expression. In Fig. 2c we show squeezing level evaluated for the optimum detection time and frequency as a function of the LOF phase for different values of $d$ (the inset shows the optimum noise frequency as a function of $\varphi$, which is independent of $d$). Note that large levels of squeezing can be found also in this non-ideal case.

## 2.5 Experimental considerations: injection of a $TEM_{10}$ seed.

In experiments, the cavity resonance is usually stabilized by using a technique called *active locking*[19,20]: the system is seeded with a low-intensity beam at the signal frequency $\omega_0$. Seeding our system with a $TEM_{10}$ mode would have an additional effect: The orientation of the bright mode would be locked to that of the seed, and hence the bright and dark modes would correspond to fixed $TEM_{10}$ and $TEM_{01}$ modes, respectively. This occurs because the injection Hamiltonian of the seed has the form

$$\hat{H}_{\text{seed}} = i\hbar\mathcal{E}_s \hat{a}^{\dagger}_{10} + \text{H.c.} = i\hbar\mathcal{E}_s (\hat{a}^{\dagger}_{+1} + \hat{a}^{\dagger}_{-1})/\sqrt{2} + \text{H.c.}, \qquad (10)$$

which does not preserve the symmetry that Hamiltonian (3) had. This should facilitate measuring the system properties.

Of course, breaking the symmetry of the system externally would destroy the phenomenon of squeezing induced by SSB, as a SSB no longer occurs. However, we thought that large squeezing levels could still be found in the $TEM_{01}$ dark mode for reasonably small levels of the seed's intensity. In order to prove this, we have studied the level of squeezing of the $TEM_{01}$ phase quadrature as a function of the seed's intensity when seeding on amplification, that is, when the seed is in-phase with the pump field and hence $\mathcal{E}_s$ is real. As will be shown elsewhere, the noise spectrum reads in this case

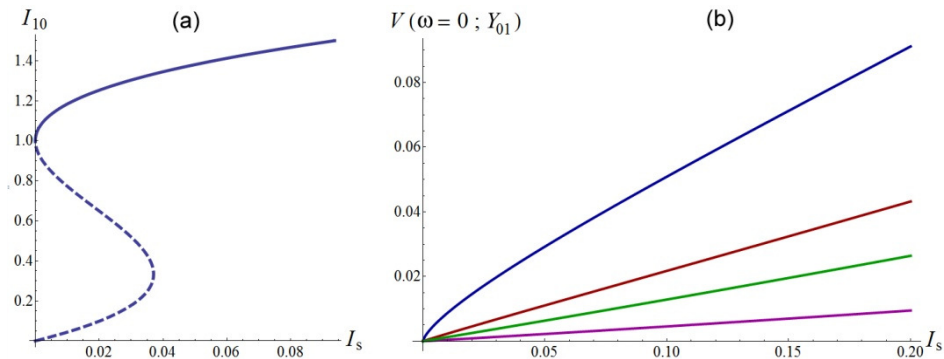

Figure 3. (a) Normalized intensity of the $TEM_{10}$ mode as a function of the injection parameter $I_s$ for $\sigma = 1.5$ (the same behavior is found for any value of $\sigma$ above threshold). Note that three solutions are allowed for small values of the injection, but dashed branches are unstable. (b) Noise spectrum of the $TEM_{01}$ mode as a function of the injection parameter $I_s$ ($\sigma = 1.001, 1.5, 2,$ and $4$, from top to bottom).

$$V(Y_{01};\omega) = 1 - 4q^2 / \left[(1+q)^2 + (\omega/\gamma_s)^2\right], \qquad (11)$$

with $q = \sigma - I_{10}$, being $I_{10}$ proportional to the intensity of the TEM$_{10}$ mode which is found from the cubic algebraic equation $[\sigma - (1 + I_{10}/2)]^2 I_{10} = I_s$. We have defined the parameter $I_s = \chi^2 \mathcal{E}_s^2 / \gamma_s^3 \gamma_p$, which can be written in terms of the external power of the seed laser $P_s$ as $I_s = 2P_s/P_{p,th}$, being $P_{p,th}$ the pump power needed to make the signal oscillate when $\mathcal{E}_s = 0$. Values of $I_s$ in the range $10^{-3}$–$10^{-1}$ are usually required for active locking. It is possible to show that only one of the three solutions of this equation is stable, namely that with the larger value (upper branch of Fig. 3a).

In Fig. 3b we show the noise spectrum at zero noise frequency (where squeezing is maximum) as a function of $I_s$ for different values of $\sigma$ (always above threshold). Notice that even for the largest seed values, squeezing is still above 90% ($V < 0.1$). For different values of the seed phase, or larger values of its intensity the system shows new interesting behavior that would be analyzed elsewhere.

## 2.6 Rotational symmetry breaking in degenerate four-wave mixing cavities

Along the previous subsections we have treated in some depth perfect and noncritical quadrature squeezing through SSB of the rotational symmetry in a special type of DOPO. In order to show that this is not specific of the considered nonlinear system, we show now that this mechanism for perfect squeezing generation can also be found in $\chi^{(3)}$ cavities.

We proposed[21] and analytically studied a special type of $\chi^{(3)}$ cavity in which the SSB of the rotational symmetry occurs in a way similar to the two-transverse-mode DOPO. The scheme of the system is depicted in Fig. 4a: Two Gaussian pumping beams of frequencies $\omega_1$ and $\omega_2$ are injected in a rotationally symmetric cavity containing an isotropic $\chi^{(3)}$ medium. The nonlinear cavity is tuned in such a way that close to the frequency $\omega_s = (\omega_1 + \omega_2)/2$ the first transverse family is resonant. One possible configuration for the cavity is explained in Fig. 4b. Within this cavity the signal field at frequency $\omega_s$ is generated through the creation of photon pairs (see the four-wave mixing process indicated in Fig. 4a), each photon having opposite OAM because of OAM conservation just as in the two-transverse-mode DOPO.

The Hamiltonian of the system is, of course, more complicated than the one treated before, and can be found in the original article[21], where the stochastic Langevin equations are also derived. The problem can be, however, simplified by treating the pumping beams as classical fields and by further assuming that they do not suffer depletion. In this limit the interaction picture Hamiltonian can be written as

$$\hat{H} = \hbar\delta\left(\hat{a}_{+1}^\dagger \hat{a}_{+1} + \hat{a}_{-1}^\dagger \hat{a}_{-1}\right) + \hbar g \left[\frac{1}{2}\left(\hat{a}_{+1}^{\dagger\,2} \hat{a}_{+1}^2 + \hat{a}_{-1}^{\dagger\,2} \hat{a}_{-1}^2\right) + 2\hat{a}_{+1}^\dagger \hat{a}_{+1} \hat{a}_{-1}^\dagger \hat{a}_{-1} + 4\rho^2\left(\hat{a}_{+1}^\dagger \hat{a}_{+1} + \hat{a}_{-1}^\dagger \hat{a}_{-1}\right) + 2\rho^2\left(\hat{a}_{+1} \hat{a}_{-1} + \hat{a}_{+1}^\dagger \hat{a}_{-1}^\dagger\right)\right], \quad (12)$$

where $\delta$ is the cavity detuning with respect to $\omega_s$, $g$ is the nonlinear coupling constant proportional to the nonlinear susceptibility of the medium[21], and $\rho$ is the amplitude of the pumping fields (which are assumed to be equal).

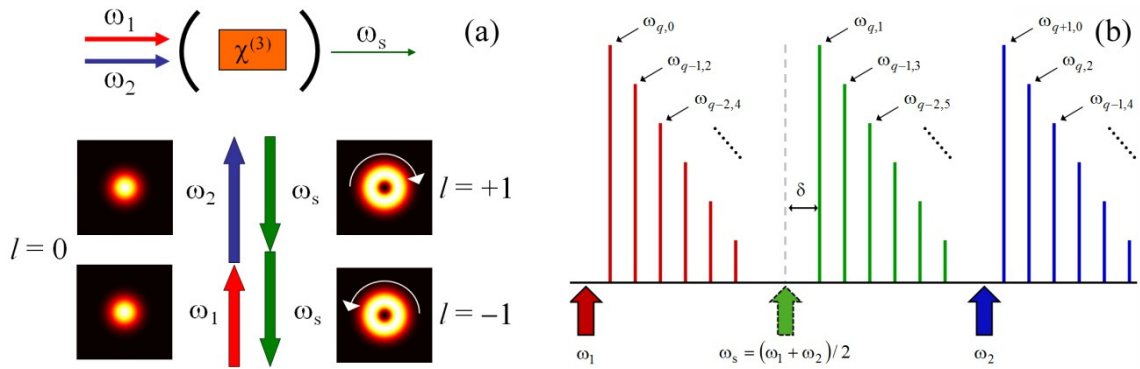

Figure 4. Scheme of a $\chi^{(3)}$ cavity for squeezing generation through SSB of the rotational symmetry. (a) While the pumping beams (frequencies $\omega_1$ and $\omega_2$) have zero OAM, the signal modes –which have frequency $\omega_s = (\omega_1 + \omega_2)/2$– have ±1 OAM (denoted as $l$ in the figure) because the cavity resonance closer to $\omega_s$ corresponds to the first family of transverse modes. (b) Resonance spectrum of a nearly symmetric, nearly confocal cavity, which is one of the possible cavity configurations with the required properties. $\omega_{q,f}$ refers to the resonance of the family transverse family of order $f$ and longitudinal mode order $q$. For this configuration one gets the maximum splitting between the $f = 1$ and $f = 0$ families.

The above Hamiltonian exhibits the symmetry $(\hat{a}_{+1},\hat{a}_{-1}) \rightarrow (e^{i\theta}\hat{a}_{+1}, e^{-i\theta}\hat{a}_{-1})$, exactly as in the two-transverse-mode DOPO treated above. Hence one can expect to obtain similar results provided that a steady, stable mean field solution for the signal modes exists in some parameter region, what is indeed proved in full detail in the original article[21]: a $TEM_{10}$ mode with an arbitrary orientation exists whenever $\delta > 3^{1/2}\gamma_s$ and the pump intensity $\rho^2$ is inside the region defined by the curves $\gamma_s/2g$ and $[2\delta + (\delta^2 - 3\gamma_s^2)^{1/2}]/6g$, and appears through a subcritical pitchfork bifurcation. The analysis of quantum fluctuations reveals that quantum noise makes this bright mode rotate, and that a quadrature of its orthogonal $TEM_{10}$ mode is perfectly squeezed within the given domain of existence of the mean field solution as expected.

## 2.7 Translational symmetry breaking in large aperture DOPOs

After the SSB of the rotational symmetry, we pass to briefly review our previous work on squeezing generation through the SSB of the translational symmetry[6,7]. In order to have translational symmetry in the transverse plane of a nonlinear cavity the resonator must have plane mirrors. Further, in order to have mean field states having a *location* in the transverse plane, a large Fresnel number cavity is needed as in this case the emission in nonlinear optical cavities tends to occur forming transverse patterns[22] (also known as dissipative structures) such as stripe or hexagonal patterns, or localized structures, such as cavity solitons. Once the bright transverse pattern $\overline{A}(\mathbf{r} - \mathbf{r}_0)$ appears in a reference location $\mathbf{r}_0$, it implies a SSB of the system's translational invariance, and hence we expect quantum noise to randomly change this location, and the appearance of noncritical and perfect squeezing in both the dark modes $i\partial_{x_0}\overline{A}(\mathbf{r} - \mathbf{r}_0)$ and $i\partial_{y_0}\overline{A}(\mathbf{r} - \mathbf{r}_0)$, with $\mathbf{r}_0 = (x_0, y_0)$.

This is what we demonstrated[6,7] with the technique showed in subsection 2.3 properly adapted to deal with the nontrivial spatial dependence of the transverse patterns. In particular, the DOPO can be modeled within the positive $P$ representation by the following stochastic equations[6,7]

$$\partial_t A_0 = \mathcal{E}_p - (\gamma_p + i\delta_p - i\gamma_p l_p^2 \nabla^2)A_0 - \frac{\chi}{2}A^2, \qquad \partial_t A_0^+ = \mathcal{E}_p - (\gamma_p - i\delta_p + i\gamma_p l_p^2 \nabla^2)A_0^+ - \frac{\chi}{2}A^{+2},$$
$$\partial_t A = -(\gamma_s + i\delta_s + -\gamma_s l_s^2 \nabla^2)A + \chi A_0 A^+ + \sqrt{\chi A_0}\,\eta(\mathbf{r},t), \qquad \partial_t A^+ = -(\gamma_s - i\delta_s + i\gamma_s l_s^2 \nabla^2)A^+ + \chi A_0^+ A + \sqrt{\chi A_0^+}\,\eta^+(\mathbf{r},t), \qquad (13)$$

where $\nabla^2$ is the transverse Laplacian, being $\delta_{p/s}$ and $l_{p/s}$ the detuning and characteristic diffraction length of the pump/signal field. The independent noises $\eta(\mathbf{r},t)$ and $\eta^+(\mathbf{r},t)$ satisfy the usual white noise statistics but now with respect to both the space and time coordinates. The complex stochastic fields $A(\mathbf{r},t)$ and $A^+(\mathbf{r},t)$ allow us to evaluate quantum averages of operator functions $f[\hat{A}(\mathbf{r},t), \hat{A}^\dagger(\mathbf{r},t)]$, as $\langle : f[\hat{A}(\mathbf{r},t), \hat{A}^\dagger(\mathbf{r},t)] : \rangle = \langle f[A(\mathbf{r},t), A^+(\mathbf{r},t)]\rangle_{\text{stochastic}}$.

The mean field equations associated to Eqs. (13) possess stationary solutions (dissipative structures) of the form $\overline{A}(\mathbf{r} - \mathbf{r}_0) = \exp[i\,\text{sign}(\delta_s)\beta]F(\mathbf{r} - \mathbf{r}_0)$, where $\beta$ is a positive constant, $F(\mathbf{r} - \mathbf{r}_0)$ is real, and $\mathbf{r}_0$ is arbitrary due to the translational invariance. In order to study the dynamics of quantum fluctuations, we adiabatically eliminate the pump field and set $A(\mathbf{r},t) = \overline{A}[\mathbf{r} - \mathbf{r}_0(t)] + b[\mathbf{r} - \mathbf{r}_0(t),t]$ and $A^+(\mathbf{r},t) = \overline{A}^*[\mathbf{r} - \mathbf{r}_0(t)] + b^+[\mathbf{r} - \mathbf{r}_0(t),t]$, where the FP of the system $\mathbf{r}_0(t)$ is allowed to vary with time. The stochastic equations are then linearized with respect to $b$, $b^+$, and $\dot{\mathbf{r}}_0$, arriving[6,7]

$$\kappa(\mathbf{v}_{0x}\dot{x}_0 + \mathbf{v}_{0y}\dot{y}_0) + \dot{\mathbf{b}} = \mathcal{L}\mathbf{b} + \boldsymbol{\eta}(\mathbf{r},t), \qquad (14.1)$$

$$\mathbf{b} = \text{col}(b,b^+), \quad \mathbf{v}_{0x(0y)} = \partial_{x(y)}\text{col}(\overline{A},\overline{A}^*), \quad \boldsymbol{\eta} = \text{col}(\sqrt{\chi A_0}\,\eta, \sqrt{\chi A_0^*}\,\eta^+), \qquad (14.2)$$

where we have defined $\kappa = \sqrt{2\gamma_s |\delta_s|}/\chi$. This equation is of the same type as (7), with a major difference: now the matrix $\mathcal{L}$, whose particular expression is not relevant for the purposes of this review and can be consulted in the original articles[6,7], is a differential operator which depends on the nontrivial transverse pattern $\overline{A}(\mathbf{r} - \mathbf{r}_0)$. This makes its full analytical diagonalization not possible; nevertheless, it is possible to show numerically that this operator always possess a biorthonormal eigensystem[6,7], that is, a set of orthogonal vectors $\{\mathbf{v}_i, \mathbf{w}_i\}$ satisfying $\{\mathcal{L}\mathbf{v}_i = \lambda_i \mathbf{v}_i, \mathcal{L}^\dagger \mathbf{w}_i = \lambda_i^* \mathbf{w}_i\}$. In particular, it is straightforward to prove that the vectors $\mathbf{v}_{0x}$ and $\mathbf{v}_{0y}$ are two Goldstone modes having null eigenvalues, while the vectors $\mathbf{w}_{1x(1y)} = i\partial_{x(y)}\text{col}(\overline{A}, -\overline{A}^*)$ have both $-2\gamma_s$ eigenvalue.

Projecting the linearized equations onto the Goldstone modes, it can be proved that the position of the transverse pattern is drifted by the quantum noise and, in particular, it has a variance given by[6,7]

$$V(\mathbf{r}_0) = \langle \delta\mathbf{r}_0^2 \rangle_{\text{stochastic}} = Dt, \text{ with } D = 2\chi\kappa^{-2}\,\text{Re}\int d^2\mathbf{r}\,(\mathbf{w}_{0x}^2 + \mathbf{w}_{0y}^2)\overline{A}_0^*, \qquad (15)$$

which increases linearly with time as the orientation of the TEM modes in the previous examples. On the other hand, projecting onto the modes $\mathbf{w}_{1x(1y)}$ it is proved[6,7] that an homodyne detection scheme in which the LOF coincides with $i\partial_{x_0} A(\mathbf{r}-\mathbf{r}_0)$ or $i\partial_{y_0} A(\mathbf{r}-\mathbf{r}_0)$ –or any linear combination of these– would lead to the squeezing spectrum (8.2), that is, this fields have perfect, noncritical squeezing on its phase quadrature. Note that when the bright field has the form of a bright cavity soliton of the sech-type, these modes are similar to a $TEM_{10}$ mode[7], while if mean emission takes place in a stripe pattern, they correspond to the stripe pattern complementary to the bright one. In any case, it was proved for the case of the cavity soliton that even when using a homogeneous LOF large levels of squeezing can be obtained[7].

### 3. POLARIZATION SYMMETRY BREAKING

In this section we generalize the idea of squeezing generation through the SSB of spatial symmetries to cases of SSB of the signal field polarization symmetry. In order to achieve non-critical quadrature squeezing through the latter, the nonlinear cavity must be invariant under variations of a FP in the polarization[23] of the signal field. This FP might be, e.g., the eccentricity or the orientation of the polarization ellipse. Notice that continuous variations of one of these parameters correspond to rotations on the Poincaré sphere defined by the Stokes parameters

In this SSB, the polarization state of the bright mode gets completely undetermined in the long term, while the dark mode coincides with the mode having orthogonal polarization with respect to the generated one. Next we resume the results found in our most recent communication[24], where we show that polarization symmetry breaking can happen, at least in principle, in both $\chi^{(2)}$ and $\chi^{(3)}$ nonlinear cavities.

#### 3.1 Polarization symmetry breaking in type-II DOPOs

Consider a type-II frequency degenerate OPO. In this device signal the photon-pairs are generated with orthogonal linear polarizations (say $\mathbf{e}_x$ and $\mathbf{e}_y$) and an undefined relative phase. This implies that the signal field has elliptical polarization (with the ellipse axes orientated at $\pm\pi/4$ with respect to the **x**-axis) but with undefined eccentricity and direction of rotation[23]. Moreover, the Hamiltonian of this system is isomorphic to that of the two-transverse-mode DOPO[8,9], Eq. (3). Hence all the conclusions we have obtained in Sects. 2.2–2.5 apply to this case, the only change being the physical meaning of the result: Now the bright mode is an elliptically polarized mode whose eccentricity will be completely undetermined in the long term, and the dark mode showing perfect squeezing is the mode with orthogonal polarization.

On the other hand, type-II OPOs being simultaneously polarization invariant and frequency degenerated do not seem to exist. In usual type-II OPOs the signal modes have different frequencies (although the frequency difference between them[25] can be as small as 150kHz). Certainly they can be made frequency degenerate, but the technique used for that purpose breaks the polarization symmetry[26]: a birefringent plate is introduced within the cavity in order to couple the two orthogonally polarized signal modes, which forces frequency degeneracy but fixes the phase difference between them thus breaking the system's polarization symmetry. Hence, given the difficulties of having frequency-degenerate type-II OPOs, we give an alternative $\chi^{(3)}$ nonlinear cavity.

#### 3.2 Polarization symmetry breaking in degenerate four-wave mixing cavities

Consider a $\chi^{(3)}$ nonlinear cavity similar to that presented in Sect. 2.6: Two Gaussian pumping beams of frequencies $\omega_1$ and $\omega_2$ *with orthogonal polarizations* are injected in a rotationally symmetric cavity containing an isotropic $\chi^{(3)}$ medium, the nonlinear cavity being tuned in such a way that close to $\omega_s = (\omega_1 + \omega_2)/2$ there is a cavity resonance. Within this cavity the signal photons are generated in pairs via four-wave mixing (see again Fig. 3).

We will not give here the Hamiltonian for this system because its expression is too lengthy[24]. Here it suffices to say that for isotropic $\chi^{(3)}$ media in which the Kleinman symmetry applies (like, e.g., for nonlinearities due to nonresonant electronic response), and for circularly polarized pump beams (as otherwise preferred transverse directions are defined), the system's Hamiltonian is isomorphic to (12), just changing the opposite OAM by opposite circular polarizations $R$ and $L$. Hence all the conclusions given in 2.6 apply to this case, in which the bright and dark mode correspond to a linearly polarized mode and its orthogonal, being their orientation the FP of the system. This $\chi^{(3)}$ cavity has the advantage over the type-II frequency degenerate OPO that it can be implemented within the experimental state of the art[27].

### 4. THE SINGLE-PHOTON-PAIR LIMIT

So far we have considered nonlinear systems in which symmetry breaking happens at a macroscopic level, as the mean field which breaks the symmetry gets highly populated. In order to understand the phenomenon from a microscopic point of view, we develop now a model in which photon-pairs are generated one at a time.

The basic scheme is depicted in Fig. 5a. A three-level atom is introduced in a cavity having highly reflective mirrors, so that losses can be neglected, focusing then on the intracavity dynamics. The atomic levels are disposed in a cascade configuration, with the additional assumption that the upper and lower levels corresponds to $J = 0$ states, while the intermediate level has $J = 1$. We assume that the cavity is tuned so that a longitudinal mode with frequency $\omega_0 = \omega_{eg}/2$ exists, where $\omega_{eg}$ is the frequency of the transition from the excited to the ground state. In the interaction picture, and further assuming that the intermediate level is detuned from the cavity resonance, and hence can be adiabatically eliminated, the Hamiltonian reads

$$\hat{H} = \hbar\chi\left(\hat{\sigma}_{g-e}\hat{a}_R\hat{a}_L + \hat{\sigma}_{e-g}\hat{a}_R^\dagger\hat{a}_L^\dagger\right), \qquad (16)$$

where $\hat{\sigma}_{g-e}$ and $\hat{\sigma}_{e-g}$ are the raising and lowering operators connecting the ground and excited states of the atom, and $\hat{a}_R$ and $\hat{a}_L$ are the annihilation operators for right and left circularly polarized photons, respectively. This Hamiltonian has the same symmetry as the ones we have worked with so far. The difference now, is that the atomic operator ensures that no more than one photon-pair is created or annihilated at the same time, as when the atom decays it must reabsorbed a photon-pair before decaying again. This Hamiltonian can be analytically diagonalized, and hence the state of the system at any time can be known for any initial state. However, in this review we will focus on the case of having the atom initially excited, and a Fock state with the same number of $R$ and $L$ photons, as this also makes the evaluation of the expressions to come analytical or semi-analytical. We leave for future works the case of starting out of a different state for the field, or a superposition of excited and ground states for the atom.

It is straightforward to show that the initial state $|\psi(0)\rangle = |e, N, N\rangle$ evolves as

$$|\psi(t)\rangle = \cos(\Omega_N t)|e, N, N\rangle - i\sin(\Omega_N t)|f, N+1, N+1\rangle, \qquad (17)$$

where $\Omega_N = \chi(N+1)$. This state shows Rabi oscillations with frequency $\Omega_N$ between the field states having $2N$ and $2(N+1)$ signal photons. We will use this state vector to evaluate the expected value of different operators.

Based on the analysis of SSB we have been explaining in the previous sections, we could expect noise reduction in the polarization mode orthogonal to the generated one. However, here the definition of the bright and dark modes is not as clear as in the previous examples, because we don't have a macroscopic field to rely on (as $\langle\hat{a}_R\rangle = \langle\hat{a}_L\rangle = 0$). Nevertheless, exploiting the analogy with the previous systems, we define the bright and dark modes by

$$\mathbf{e}_b = \frac{1}{\sqrt{2}}\left(e^{-i\hat\theta}\mathbf{e}_R + e^{i\hat\theta}\mathbf{e}_L\right), \quad \mathbf{e}_d = -\frac{i}{\sqrt{2}}\left(e^{-i\hat\theta}\mathbf{e}_R - e^{i\hat\theta}\mathbf{e}_L\right), \qquad (18)$$

where now $\theta$, which is half the phase-difference between the signal modes, is not a stochastic variable (as we are not using a coherent representation), but a fully quantum-mechanical operator. As it is well known, a satisfactory phase operator of a single light mode has not been found yet. Nevertheless, even though not without debate, a phase difference operator satisfying reasonable properties was found by Luis and Sánchez-Soto[28]. In this reference the eigenvectors associated to the phase-difference operator are expressed as a function of the two-mode number state basis, and, using the spectral theorem, this allows us to express any function of the operator $\hat\theta$ as

$$f(\hat\theta) = \sum_{n=0}^{\infty}\frac{1}{n+1}\sum_{r,m,m'=0}^{n} f\left(-\frac{\phi_r^{(n)}}{2}\right)e^{i(m-m')\phi_r^{(n)}}|m, n-m\rangle\langle m', n-m'|, \quad \phi_r^{(n)} = \phi_0 + \frac{2\pi r}{n+1}, \qquad (19)$$

being $\phi_0$ an arbitrary phase which has the meaning of the vacuum modes phase difference.

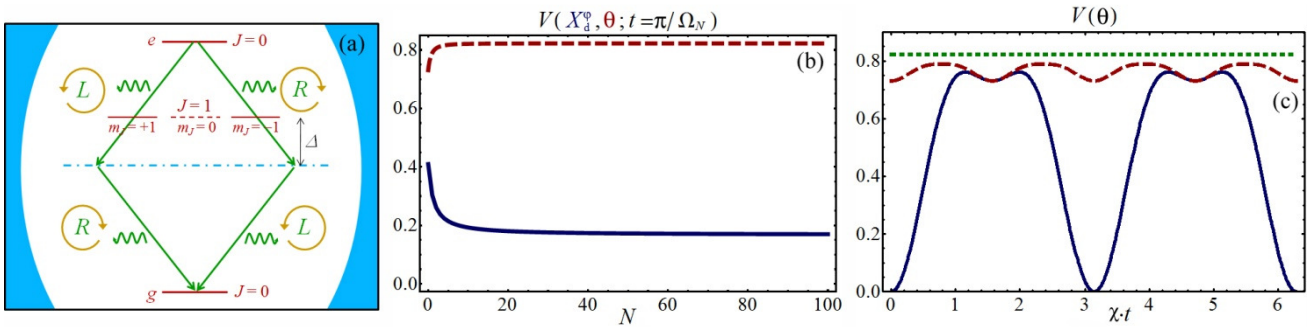

Figure 5. (a) An atom with two $J = 0$ levels plus an intermediate $J = 1$ level is introduced in a high Q cavity. Electric-dipole transitions require $|\Delta J|=0$, and hence when the atom decays from the excited to the ground state, the generated cavity photons must have orthogonal circular polarizations. (b) Variances of the phase $\theta$ (red-dashed) and the dark quadratures (blue-solid) at the middle of a Rabi oscillation as a function of the initial number of pairs $R$–$L$. (c) Variance of the phase q as a function of time for $N = 0$, 1, and 100 (blue-solid, red-dashed, and green-dotted, respectively). Note that as $N$ increases the amplitude of the oscillations decreases.

We write the boson modes associated to the bright and dark modes as

$$\hat{a}_b = \frac{1}{\sqrt{2}}\left(e^{i\hat{\theta}}\hat{a}_R + e^{-i\hat{\theta}}\hat{a}_L\right), \quad \hat{a}_d = \frac{i}{\sqrt{2}}\left(e^{i\hat{\theta}}\hat{a}_R - e^{-i\hat{\theta}}\hat{a}_L\right), \quad (20)$$

where the order of the exponentials is chosen so that $\hat{a}_b^\dagger\hat{a}_b + \hat{a}_d^\dagger\hat{a}_d = \hat{a}_R^\dagger\hat{a}_R + \hat{a}_L^\dagger\hat{a}_L \equiv \hat{N}_s$, that is, the number operator for the photons is left unchanged. Note that although for state (17) the mean field is zero for both modes, the mean number of dark photons is zero $-\langle\hat{a}_d^\dagger\hat{a}_d\rangle = 0-$, while the mean number of bright photons coincides with the total mean number of signal photons $-\langle\hat{a}_b^\dagger\hat{a}_b\rangle = \langle\hat{N}_s(t)\rangle = 2N + 2\sin^2(\Omega_N t)-$. This makes the interpretation of the modes as a bright and a dark mode somehow appropriate.

As for the squeezing levels of the dark mode, it is easy to evaluate the variance of any of its quadratures $\hat{X}_d^\varphi$; we get

$$V(X_d^\varphi) = \langle(\delta\hat{X}_d^\varphi)^2\rangle = 1 + \left[\frac{2N^2}{2N+1} - s(N)\right]\cos^2(\Omega_N t) + \left[\frac{2(N+1)^2}{2N+3} - s(N+1)\right]\sin^2(\Omega_N t), \quad (21.1)$$

$$s(M) = \frac{1}{(2M+1)^2}\sum_{m=0}^{2M}\sqrt{m(2M+1-m)}\sin^{-2}\left[\frac{(M-m+1/2)\pi}{2M+1}\right]. \quad (21.2)$$

Note that the variance is independent of φ, what makes sense as starting out of number states (which doesn't have a definite phase) should not privilege any specific phase of the signal field. This expression shows that squeezing is maximum at the middle of the Rabi oscillations and increases with $N$. In Fig. 5b we show the minimum of the variance as a function of $N$; an asymptotic value of 0.17 is obtained, and hence in this case intracavity squeezing is not perfect.

In order to show that the squeezing of the dark mode is related to the indetermination of the operator q, we can also evaluate the variance of this; a straightforward calculation shows that

$$V(\theta) = \langle(\delta\hat{\theta})^2\rangle = \frac{\pi^2}{3}\left\{\frac{N(N+1)}{(2N+1)^2} + \frac{(5+2N)\sin^2(\Omega_N t) - 3\sin^4(\Omega_N t)}{[3+4N(2+N)]^2}\right\}. \quad (22)$$

In Fig.5c we show this variance as a function of time. Note that it is maximum at the middle of the Rabi oscillations (except for $N=0$). In Fig. 5b we also show this variance at the middle of the Rabi oscillation as a function of $N$, showing that it arrives to an asymptotic value of $\pi^2/12$. Note that the behavior of the phase variance is then completely opposite to that of the dark mode quadratures: when the latter is better defined (more squeezed), the former gets more undetermined. This is an indirect proof of the connection between these two phenomena. It also explains why squeezing in the dark mode is not perfect in this case: it cannot be perfectly determined just because θ never gets completely undetermined.

## 5. CONCLUSIONS

In this paper we have reviewed the work done by our group up to date on the subject of generation of perfect and non-critical quadrature squeezing through spontaneous symmetry breaking (SSB), which includes spatial symmetries (both rotational and translational) as well as polarization symmetry. We have presented results on the squeezing properties of several nonlinear cavities, both of $\chi^{(2)}$ and $\chi^{(3)}$ types, in which these phenomena occur as well as discussed some aspects concerning the possibilities of observing them. Of particular relevance for the measurability of the squeezing achieved by these systems is the fact that the free parameter of the mean field diffuses: the orientation of the bright mode (in the case of rotational symmetry), the location of the bright transverse pattern (in the case of translational symmetry), and the orientation or eccentricity of the polarization of the bright mode (in the case of polarization symmetry) vary randomly in time, as they are governed by Wiener processes. This makes impossible to match the local oscillator to the output field in a homodyne detection scheme. However, the fact that the bright mode has a huge number of photons unless the system operates too close to threshold implies that the diffusion is quite slow, and the results given in Sect. 2.5 suggest that the detected squeezing level can be very large within experimentally achievable conditions. Moreover, as discussed in Sect. 2.6, with a slight deliberate breaking of the underlying symmetry still large squeezing levels are produced while at the same time diffusion is prevented. Finally, in the last section we have proposed a cavity QED scheme in which SSB can be investigated at the very fundamental level. We showed that the relation between the indetermination of the free parameter and the squeezing of the dark mode –which was previously proved for macroscopic fields– appears already when a single photon-pair is generated.

This work has been supported by the Spanish Ministerio de Educación y Ciencia and the European Union FEDER through Project FIS2008-06024-C03-01. C.N.-B. acknowledges a grant from the FPU Programme of the Spanish Government.